\documentclass[12pt]{article}
\usepackage[pdftex]{graphicx}

\usepackage{epsfig,amsmath,amssymb,verbatim,mathrsfs,subfigure}

\def\beq{\begin{eqnarray}}
\def\eeq{\end{eqnarray}}
\def\bea{\begin{eqnarray}}
\def\eea{\end{eqnarray}}

\def\tev{\, {\rm TeV}}
\def\gev{\, {\rm GeV}}

\newcommand{\gsim}{\lower.7ex\hbox{$\;\stackrel{\textstyle>}{\sim}\;$}}
\newcommand{\lsim}{\lower.7ex\hbox{$\;\stackrel{\textstyle<}{\sim}\;$}}

\def\stilde{\widetilde}

\newcommand{\newc}{\newcommand}
\newc{\Nc}{N_{c}}
\newc{\CG}{C_G}
\newc{\gp}{g'}
\newc{\stopi}{\stilde t_i}
\newc{\sboti}{\stilde b_i}
\newc{\staui}{\stilde \tau_i}
\newc{\stopj}{\stilde t_j}
\newc{\sbotj}{\stilde b_j}
\newc{\stauj}{\stilde \tau_j}
\newc{\stopI}{\stilde t_1}
\newc{\stopII}{\stilde t_2}
\newc{\sbotI}{\stilde b_1}
\newc{\sbotII}{\stilde b_2}
\newc{\stauI}{\stilde \tau_1}
\newc{\stauII}{\stilde \tau_2}
\newc{\sstop}{s_{t}}
\newc{\cstop}{c_{t}}
\newc{\ssbot}{s_{b}}
\newc{\csbot}{c_{b}}
\newc{\sstau}{s_{\tau}}
\newc{\cstau}{c_{\tau}}
\newc{\Sstop}{s_{2t}}
\newc{\Cstop}{c_{2t}}
\newc{\Ssbot}{s_{2b}}
\newc{\Csbot}{c_{2b}}
\newc{\Sstau}{s_{2\tau}}
\newc{\Cstau}{c_{2\tau}}
\newc{\salpha}{s_\alpha}
\newc{\calpha}{c_\alpha}
\newc{\Calpha}{c_{2\alpha}}
\newc{\Salpha}{s_{2\alpha}}
\newc{\sbetapm}{s_{\beta_\pm}}
\newc{\cbetapm}{c_{\beta_\pm}}
\newc{\Sbetapm}{s_{2 \beta_\pm}}
\newc{\Cbetapm}{c_{2 \beta_\pm}}
\newc{\sbetaO}{s_{\beta_0}}
\newc{\cbetaO}{c_{\beta_0}}
\newc{\SbetaO}{s_{2 \beta_0}}
\newc{\CbetaO}{c_{2 \beta_0}}
\newc{\vu}{v_u}
\newc{\vd}{v_d}
\newc{\seL}{\stilde e_L}
\newc{\smuL}{\stilde \mu_L}
\newc{\seR}{\stilde e_R}
\newc{\smuR}{\stilde \mu_R}
\newc{\suL}{\stilde u_L}
\newc{\sdL}{\stilde d_L}
\newc{\suR}{\stilde u_R}
\newc{\sdR}{\stilde d_R}
\newc{\scL}{\stilde c_L}
\newc{\ssL}{\stilde s_L}
\newc{\scR}{\stilde c_R}
\newc{\ssR}{\stilde s_R}
\newc{\snue}{\stilde \nu_e}
\newc{\snumu}{\stilde \nu_\mu}
\newc{\snutau}{\stilde \nu_\tau}
\newc{\Gpm}{G^\pm}
\newc{\Hpm}{H^\pm}
\newc{\FFbS}{\overline{FF}S}
\newc{\FFbV}{\overline{FF}V}
\newc{\FSS}{F_{SS}}
\newc{\FSSS}{F_{SSS}}
\newc{\FFFS}{F_{FFS}}
\newc{\FFFbS}{F_{\overline{FF}S}}
\newc{\FSSV}{F_{SSV}}
\newc{\FVS}{F_{VS}}
\newc{\FVVS}{F_{VVS}}
\newc{\FFFV}{F_{FFV}}
\newc{\FFFbV}{F_{\overline{FF}V}}
\newc{\Fgauge}{F_{\rm gauge}}
\newc{\DRbarprime}{$\overline{\rm DR}'$ }
\newc{\DRbar}{$\overline{\rm DR}$ }
\newc{\MSbar}{$\overline{\rm MS}$ }
\newc{\Yu}{{\bf Y}_u}
\newc{\Yd}{{\bf Y}_d}
\newc{\Ye}{{\bf Y}_e}
\newc{\Au}{{\bf a}_u}
\newc{\Ad}{{\bf a}_d}
\newc{\Ae}{{\bf a}_e}
\newc{\bm}{{\bf m}}
\newc{\zhol}{Z^{\rm hol}}

\newc{\rwino}{r_{\tilde W}}
\newc{\rmu}{r_{\tilde H}}
\newc{\ra}{r_A}
\newc{\ccdot}{\!\cdot\!}

\begin{document}

\setlength{\baselineskip}{0.2in}



\begin{titlepage}
\noindent
\begin{flushright}
{\small CERN-PH-TH/2008-071} \\
{\small MCTP-08-13}  \\
{\small KIAS-P08025}\\
{\small NSF-KITP-08-18}
\end{flushright}
\vspace{-.5cm}

\begin{center}
  \begin{Large}
    \begin{bf}
Higgs Boson Exempt No-Scale Supersymmetry\\
with a Neutrino Seesaw:\\
Implications for Lepton Flavor Violation\\
and Leptogenesis\\
     \end{bf}
  \end{Large}
\end{center}
\vspace{0.2cm}
\begin{center}
\begin{large}
Eung~Jin~Chun$^a$, Jason~L.~Evans$^{b}$,
David~E.~Morrissey$^{b,d}$,  James~D.~Wells$^{c,b,d}$ \\
\end{large}
  \vspace{0.3cm}
  \begin{it}
$^a$~Korean Institute for Advanced Study (KIAS) \\
Hoegiro 87, Dongdaemun-gu Seoul 130-722, Korea
\vspace{0.2cm}\\
$^b$~Michigan Center for Theoretical Physics (MCTP) \\
University of Michigan, Ann Arbor, MI 48109
\vspace{0.2cm}\\
$^c$~CERN, Theory Division, CH-1211 Geneva 23, Switzerland
\vspace{0.2cm}\\
$^d$~Kavli Institute for Theoretical Physics \\
University of California, Santa Barbara, CA 93106-4030

\vspace{0.1cm}
\end{it}

\end{center}

\center{\today}

\begin{abstract}

  Motivated by the observation of neutrino oscillations, we extend
the Higgs boson exempt no-scale supersymmetry model (HENS) by
adding three heavy right-handed neutrino chiral supermultiplets to
generate the light neutrino masses and mixings. The neutrino
Yukawa couplings can induce new lepton flavor violating couplings
among the soft terms in the course of renormalization group
running down from the boundary scale. We study the effects this
has on the predictions for low-energy probes of lepton flavor
violation~(LFV). Heavy right-handed neutrinos also provide a way
to generate the baryon asymmetry through leptogenesis. We find
that consistency with LFV and leptogenesis puts strong
requirements on either the form of the Yukawa mass matrix or the
smallness of the Higgs up soft mass. In all cases, we generically
expect that new physics LFV is non-zero and can be found in a
future experiment.

\end{abstract}

\vspace{1cm}

\end{titlepage}

\setcounter{page}{2}

\tableofcontents



\section{Introduction}

  Supersymmetry~(SUSY) is a well-motivated and elegant possibility
for new physics beyond the Standard Model~(SM).  However, SUSY can
only be an approximate symmetry of nature.  The requirement of
(soft) SUSY breaking introduces many new unconstrained parameters
to the theory that can be phenomenologically problematic.
For example, generic soft supersymmetry breaking couplings
of $\tev$ size would lead to excessive amounts of flavor mixing
and CP violation~\cite{Hagelin:1992tc,Gabbiani:1996hi}.

  A simple way to address this flavor-mixing problem of
low-energy supersymmetry is to arrange for all the matter scalar
soft terms to vanish at a common input scale $M_c$. Provided this
input scale is well above the electroweak scale, acceptably large
scalar soft terms will be regenerated in the course of
renormalization group running from the high scale to the scale of
the soft supersymmetry breaking couplings~\cite{Ellis:1984bm,
Kaplan:1999ac,Chacko:1999mi,Schmaltz:2000gy,Nelson:2000sn}.
Since the scalar soft terms generated in this way come mostly from
loops of gauginos, they are nearly flavor universal and therefore
consistent with the current bounds on flavor mixing.

  This scenario for addressing the SUSY flavor problem is realized
within Higgs-exempt no-scale supersymmetry~(HENS)~\cite{Evans:2006sj}.
In this model, the squark and slepton soft terms
all vanish at a high input scale, taken to be the scale of
unification $M_{GUT}\simeq 2\times 10^{16}\,\gev$, while the
gaugino masses are non-zero there.  This can be achieved within an
extra-dimensional setup as in gaugino mediation~\cite{
Kaplan:1999ac,Chacko:1999mi,Schmaltz:2000gy}, or by nearly conformal
running~\cite{Nelson:2000sn,Luty:2001jh,Roy:2007nz,Murayama:2007ge}.
However, unlike pure gaugino
mediation and traditional no-scale models, the Higgs scalar
squared masses are allowed to be non-vanishing at the input scale
$M_{GUT}$.  With this small modification, that does not contribute appreciably
to flavor mixing, it is possible to obtain a cosmologically-favored
neutralino LSP~\cite{Evans:2006sj}.
Under the assumption of gaugino universality, the free parameters
of the HENS model at the input scale $M_{GUT}$ are
\beq
\tan\beta,~m_{H_u}^2,~m_{H_d}^2,~M_{1/2},~sgn(\mu),
\eeq
where $M_{1/2}$ is the universal gaugino mass.
With this small number of inputs, the HENS model is able to
account for the dark matter, can be made consistent with all
current experimental bounds, and leads to exciting collider
phenomenology~\cite{Evans:2006sj}.

  While the HENS model is phenomenologically enticing,
it cannot explain the observation of neutrino
oscillations~\cite{Mohapatra:2005wg}.
This shortcoming can be resolved by supplementing the
model with three heavy singlet right-handed neutrino chiral
superfields with the superpotential couplings
\beq W=W_0+NY_\nu L H_u +
\frac{1}{2}NM_NN,\label{SupPo}
\eeq
where $W_0$ is the MSSM superpotential, $N$ are the right-handed neutrinos,
$M_N$ is their Majorana mass matrix, and $Y_\nu$ is the neutrino Yukawa matrix.
By taking the singlet neutrino masses $M_{N_i}$ to be much larger
than the electroweak scale, very small masses can be generated for
the left-handed neutrinos by the seesaw mechanism~\cite{Mohapatra:2005wg}.
Integrating out the heavy neutrino states yields the
effective superpotential coupling
\beq
W_{eff}=W_0-\frac{1}{2}(Y_{\nu}^TM_N^{-1}Y_{\nu})_{ij}(L_iH_u)(L_jH_u).
\label{Weff}
\eeq
For $M_{N} \sim 10^{12}\,\gev$, this interaction can generate
correct light neutrino masses at the weak scale with the neutrino
Yukawa couplings on the order of unity, $Y_{\nu}\sim 0.1$.

  Adding heavy right-handed neutrinos to the HENS scenario also
introduces a new flavor-mixing problem to the model. In running
the soft parameters in the full theory (Eq.~\eqref{SupPo})
from the input scale $M_{GUT}$ down to the heavy singlet neutrino
scale $M_N$, the neutrino Yukawa couplings generate non-universal
contributions to the soft masses for the charged
leptons~\cite{Borzumati:1986qx}. Such couplings are dangerous
because they are a source of lepton flavor
violation~(LFV)~\cite{Borzumati:1986qx}, for which the experimental
bounds are extremely strong.  This in turn imposes stringent
constraints on the heavy neutrino sector.

  Although adding right-handed neutrinos to SUSY models can
lead to problematic LFV rates, such extensions also have some
attractive collateral features. One of these is the possibility
of generating the baryon asymmetry via
leptogenesis~\cite{Fukugita:1986hr,Luty:1992un}.
Heavy right-handed neutrinos provide all the necessary
ingredients for baryogenesis.
Lepton number is not a conserved quantity in the neutrino sector
since the Majorana masses of the heavy right-handed neutrinos
violate lepton number $L$ by two units.
Combined with the $(B\!+\!L)$-violation due to $SU(2)_L$ sphaleron
transitions in the early universe~\cite{Klinkhamer:1984di,Kuzmin:1985mm},
there exists a source of baryon number violation.
The neutrino sector also provides
a new source of CP violation from the complex neutrino Yukawa matrix.
This CP violation can manifest itself in the out-of-equilibrium decays
and scatterings of the right-handed neutrinos in the early universe.
Together, these features fulfill the three Sakharov conditions
for baryogenesis~\cite{Sakharov:1967dj}, which can be realized
through the mechanism of leptogenesis.

  Requiring that the neutrino-extended HENS ($\nu$HENS) model account
for the baryon asymmetry of the universe while respecting the
current bounds on LFV leads to constraints on the structure of the
neutrino Yukawa matrix and the right-handed neutrino masses.
Previous studies combining the requirements for leptogenesis with
the bounds from LFV can be found in Refs.~\cite{Pascoli:2003rq,
CH04,Petcov:2005jh,Antusch:2006vw}. Compared to these previous
works, we study the constraints from LFV within the context of a
specific model for which the lack of flavor mixing outside the
neutrino sector is well-motivated. An interesting result along
these lines is that the amount of LFV in the HENS model is largely
controlled by the value of $m_{H_u}^2$ at the high input scale.
The degree to which the neutrino sector parameters are constrained
therefore depends strongly on the size of $m_{H_u}^2$.

  The outline of this paper is as follows.  In Section~\ref{LFVHN}
we investigate LFV in the HENS model induced by the inclusion
of heavy right-handed neutrinos.  Using the current bounds on
LFV processes, we obtain constraints on the underlying model.
In Section~\ref{TheLep} we investigate whether it is possible
for the HENS model with right-handed neutrinos to account for
the baryon asymmetry by way of thermal leptogenesis.
We combine the results of Sections~\ref{LFVHN} and \ref{TheLep}
in Section~\ref{LepLFV}, where we examine whether leptogenesis
can generate the baryon asymmetry while satisfying bounds from LFV.
Finally, Section~\ref{concl} is reserved for our conclusions.

\section{LFV in the HENS Model with Heavy Neutrinos\label{LFVHN}}

  We begin by considering the constraints on the HENS model
from LFV induced by the inclusion of heavy right-handed
neutrinos.  These constraints depend strongly on the parameters
in the neutrino sector such as the Majorana masses
for the right-handed neutrinos and the neutrino
Yukawa matrix.  Some of these neutrino sector parameters have
been determined by the measurements of the light neutrino mass
differences and mixings~\cite{Maltoni:2004ei,Fogli:2005cq}.
In anticipation of computing the LFV constraints, we collect
here our notation and assumptions about the neutrino sector.

  In terms of the couplings in the full superpotential of
Eq.~\eqref{SupPo}, the low-energy effective superpotential of
Eq.~\eqref{Weff} implies that the light neutrino mass matrix is
given by
\beq
m_{\nu_{ij}}=\frac{v_u^2}{2}(Y_{\nu}^TM_N^{-1}Y_{\nu})_{ij}
\eeq
This matrix can be diagonalized by the unitary PMNS matrix
$U$~\cite{Pontecorvo:1957cp,Maki:1962mu}. Following the standard
convention, we will parameterize the PMNS matrix with three real
angles and three phases according to
\begin{equation}
U=
\mathcal{O}_{23}(\theta_{23})\,\Gamma_\delta
\mathcal{O}_{13}(\theta_{13})\,\Gamma_\delta^*\,
\mathcal{O}_{12}(\theta_{12})\,\times
diag[e^{i\alpha_1/2},e^{i\alpha_2/2},1]\label{Umat}
\end{equation}
where $\Gamma_\delta =diag(1,1,e^{i\delta})$, and
$\mathcal{O}_{ij} = [(c_{ij}, s_{ij}),(-s_{ij}, c_{ij})]$ with
$c_{ij}=\cos\theta_{ij}$ and $s_{ij}=\sin\theta_{ij}$.

  It is convenient to make use of the known structure of the light
neutrino mass matrix to parameterize the neutrino Yukawa matrix
$Y_{\nu}$ according to~\cite{Casas:2001sr}
\begin{equation}
Y_\nu=\frac{1}{v_u}\,\sqrt{M_N}\,R\,
\sqrt{m_{\nu_{diag}}}\,U^{\dagger}\label{NueYuka}
\end{equation}
where $R$ is a \emph{complex orthogonal}
matrix, $M_N$ is the diagonal right-handed
neutrino mass matrix, and $m_{\nu_{diag}}$ is the diagonalized
left-handed neutrino mass matrix.  Here, and throughout this paper,
we will always work in a field basis such that the right-handed neutrino
and charged lepton mass matrices are diagonal.
Since the $R$ matrix is complex orthogonal, we can parameterize it in
terms of three \emph{complex} angles according to
\begin{equation}
R=
diag(\pm1,\pm1,\pm1)\;
\mathcal{O}_{12}(\theta_{12R})\,\mathcal{O}_{23}(\theta_{23R})\,
\mathcal{O}_{31}(\theta_{31R}).\label{Rmat}
\end{equation}
with $\mathcal{O}_{ij} = [(c_{ijR},s_{ijR}),(-s_{ijR}, c_{ijR})]$, where
$c_{ijR}=\cos\theta_{ijR}$ and $s_{ijR}=\sin\theta_{ijR}$. Note
that since these angles are complex, the components of $R$ are not
bounded in magnitude.  This means that some of the entries in the
neutrino Yukawa matrix could be quite large,
but through cancellations among the see-saw contributions,
still give rise to acceptably small light neutrino masses.
In order to avoid too much fine-tuning in this regard, we will only
consider $R$ matrices with $|R_{ij}|<10$, which corresponds
roughly to a tuning of less than $10\%$ in the light neutrino mass
matrix.  Our choices for the light neutrino masses and mixings
are listed in Appendix A.

\subsection{Off-Diagonal HENS Soft Terms from RG Running\label{RGrun}}

  Without heavy right-handed neutrinos, the HENS model is
safe in terms of lepton-flavor violation (LFV).  With heavy
right-handed neutrinos, lepton-flavor violating couplings
can arise among the scalar soft terms in the course of renormalization
group~(RG) running down from the input scale $M_{GUT}$.
The strict experimental limits on LFV will in turn
lead to constraints on the neutrino Yukawa couplings and right-handed
neutrino masses.  Since this new source of FCNC in the HENS model
arises from RG running, and not the SUSY breaking mechanism, its amplitude
will have a similar form to that found in mSUGRA models.

  The dominant contribution to the off-diagonal flavor-mixing components
of the scalar soft squared masses is well-approximated by keeping
only the leading logarithmic term in the RG running.\footnote{ The
leading-log approximation breaks down for $M_{1/2}\gsim 1000\gev$
and $|m_{H_u}^2|\lesssim (100\,\gev)^2$~\cite{Petcov:2003zb}.  To
avoid this, we include subleading terms in our numerical
analysis.} With this approximation applied to the boundary
conditions appropriate to the HENS model ($m_{\tilde{f}}^2 = 0,
~m_{H_u}^2,\,m_{H_d}^2\neq 0$), we
obtain~\cite{Borzumati:1986qx,Hisano:1995nq} \beq
{m}_{\tilde{L}_{i\neq j}}^2=-\frac{1}{8\pi^2}m^2_{Hu}\sum_k Y_{\nu
ki}^* Y_{\nu kj}\, \ln\left(\frac{M_{GUT}}{M_{N_k}}\right).
\label{MasIns} \eeq To this order of approximation, the flavor
non-diagonal elements in the scalar trilinear soft couplings and
the right-handed slepton soft masses vanish. When the constraints
on the neutrino Yukawa couplings from LFV are applied, the
corrections to the diagonal components of the scalar masses are
numerically very small; less than about $5\,\gev$ in most of the
parameter space.  However, when these corrections could be
relevant we have included them.

\subsection{HENS LFV\label{HENlfv}}

  The off-diagonal soft terms introduced by RG running,
given in Eq.~\eqref{MasIns}, will induce
LFV transitions of the type $\ell_i\to \ell_j\,\gamma$.
The leading contributions to the branching fractions
for these transitions can be written as~\cite{Hisano:1995nq,
Hisano:1998fj,Masina:2002mv}
\begin{equation}
B(\ell_i\rightarrow \ell_j \gamma) =
\frac{\alpha}{4\,\Gamma(\ell_i)}\,m_{\ell_i}^5\,
|A_L^{(ij)}|^2,
\label{BraFrac}.
\end{equation}
where $\Gamma(\ell_i)$ is the total decay width of lepton
$\ell_i$, and the amplitude $A_{L}^{(ij)}$ has the schematic
form~\cite{Hisano:1995nq,Hisano:1998fj}
\beq A_{L}^{(ij)} =
{m}_{\tilde{L}_{i\neq j}}^2\, F^{(ij)}_{L},
\label{bramplitude}
\eeq
with $F^{(ij)}_{L}$ a combination of loop functions that depend
on the chargino, neutralino, and slepton masses.
These loop functions are such that the dominant contribution to
$B(\ell_i\to\ell_j\gamma)$ scales approximately as
${m}_{\tilde{L}_{i\neq j}}^2\,\tan^2\beta\,M_{1/2}^{-8}$.
Note also that in this leading contribution to the LFV branching
fractions, the flavor violating term ${m}_{\tilde{L}_{i\neq j}}^2$
can be factored out.  This will allow us to discuss the effects
of the neutrino sector and the supersymmetry breaking sector separately.

  The differences in the branching fractions of
Eq.~(\ref{BraFrac}) for the HENS model compared to mSUGRA lie in
the form ${m}_{\tilde{L}_{i\ne j}}^2$ and the low-scale sparticle masses.
However, ${m}_{\tilde{L}_{i\ne j}}^2$ is
qualitatively similar in the two theories and will be of the same
order of magnitude for both theories as long as $m_{H_u}^2\sim
m_0^2+a_0^2$.  The loop functions $F_L^{(ij)}$ are also
qualitatively similar, but differ in the masses of the gauginos
and sleptons running in the loops that appear as their arguments.
From this, there can be a slight enhancement of the LFV rates in
HENS relative to mSUGRA because the slepton masses tend to be
somewhat lighter in the HENS model.
On the other hand, the LFV rates can be reduced in the HENS
model relative to mSUGRA by arranging for $m_{H_u}^2$ to vanish,
which suppresses the leading source of lepton flavor mixing
given in Eq.~\eqref{MasIns}.  As shown in Ref.~\cite{Evans:2006sj},
it is often possible to obtain a consistent phenomenology
with $m_{H_u}^2 \sim 0$, especially for $\tan\beta\lesssim 30$.
To obtain a similar suppression in mSUGRA, one would need
both $m_0$ and $a_0$ to be quite small which can be phenomenologically
problematic~\cite{Feng:2005ba,de Austri:2006pe}.

\subsection{Constraints on the HENS Model from LFV\label{ConLFV}}

   The possibility of inducing LFV places significant
constraints on right-handed neutrino extensions of the HENS model.
The two strongest bounds on new sources of LFV come from searches
for $\mu\to e\gamma$ and $\tau\to \mu\gamma$ transitions:
\bea
B(\mu\rightarrow e\gamma) &<& 1.2\times 10^{-11},~~~~~~~~\cite{
Brooks:1999pu}
\label{lfvconstr1}\\
B(\tau\rightarrow\mu\gamma) &<& 4.5\times 10^{-8},\,~~~~~~\cite{
Aubert:2005ye,Hayasaka:2007vc}
\label{lfvconstr2}\\
B(\tau\rightarrow e\gamma) &<& 1.1\times 10^{-7},\,~~~~~~~~~\cite{
Aubert:2005wa}
\label{lfvconstr3}
\eea
It was shown in Ref.~\cite{Hisano:1995nq} that if these bounds are
satisfied, the bounds on other experimentally searched-for
channels such as $B(\mu\to 3e)$ will generally be satisfied as well.

\begin{figure}[ttt]
\begin{center}
  \includegraphics[width=0.7\textwidth]{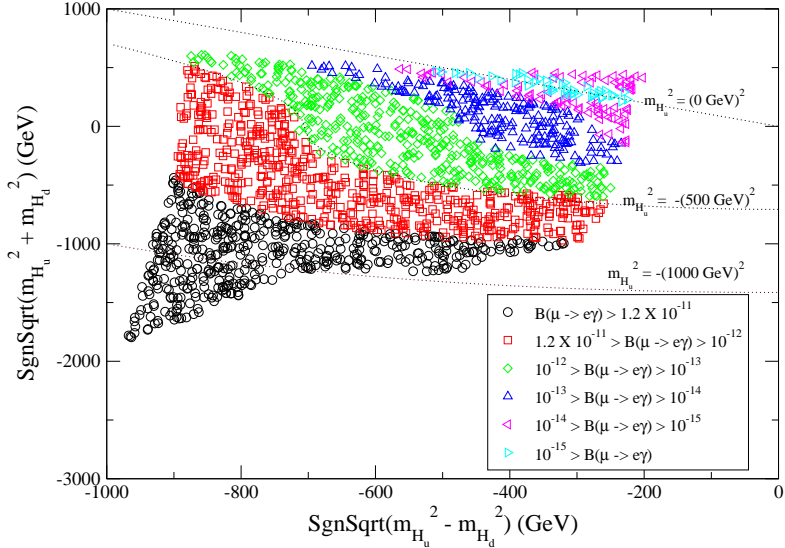}
\vspace{0.3cm} \caption{\label{BRm300b10} $B(\mu\rightarrow
e\gamma)$ as a function of the HENS model parameters $m_{H_u}^2$
and $m_{H_d}^2$ at the high input scale. The other model
parameters are $M_{1/2}=300\gev$, $\tan\beta=10$, and $sgn(\mu) = 1$
as well as neutrino-sector parameters
$\theta_{12R}=\theta_{13R}=\theta_{23R}=\pi/4+i\ln(\sqrt{2})$,
$M_{N_3}=10^{12}\,\gev$, $M_{N_2}=10^{11}\,\gev$, and
$M_{N_1}=10^{10}\,\gev$. All points in this plot are consistent with
collider phenomenology constraints and have a neutralino LSP. }
\end{center}
\end{figure}

   In Fig.~\ref{BRm300b10} we show the dependence of the LFV branching
fraction $B(\mu\to e\gamma)$ on the high-scale input values of
$m_{H_u}^2$ and $m_{H_d}^2$ in the HENS model with right-handed
neutrinos.  The other HENS parameters are taken to be $M_{1/2} =
300\,\gev$, $\tan\beta = 10$, and $sgn(\mu) = 1$.
This value of $M_{1/2}$ is about as small as is possible in the HENS
model while still obtaining a sufficiently heavy
Higgs boson~\cite{Evans:2006sj}.
The points in this figure cover the region of the HENS parameter
space that is consistent with all collider and phenomenological constraints
other than from LFV, and that has a neutralino LSP. The
neutrino-sector parameters are taken to be $M_{N_3}=10^{12}$ GeV,
$M_{N_2}=10^{11}$ GeV, $M_{N_1}=10^{10}$ GeV, the light neutrino
masses are as described in Appendix~A with $m_3 = 0.05\,\mbox{eV}$, and the
$R$-matrix angles (see Eq.~\eqref{Rmat}) are equal to
$\theta_{12R}=\theta_{13R}=\theta_{23R}=\pi/4+i\ln(\sqrt{2})$.
These particular values of the neutrino sector parameters were
chosen for convenience, but we have checked that they lead to
typical amounts of LFV.  The decreasing trend in $B(\mu\to
e\gamma)$ from bottom-left to top-right in this figure corresponds
largely to a decreasing value of $m_{H_u}^2$.  This is not surprising
given Eq.~\eqref{MasIns}, which shows that the leading contribution
to lepton flavor mixing is proportional to $m_{H_u}^2$.

\begin{figure}[ttt]
\begin{center}
\includegraphics[width=0.7\textwidth]{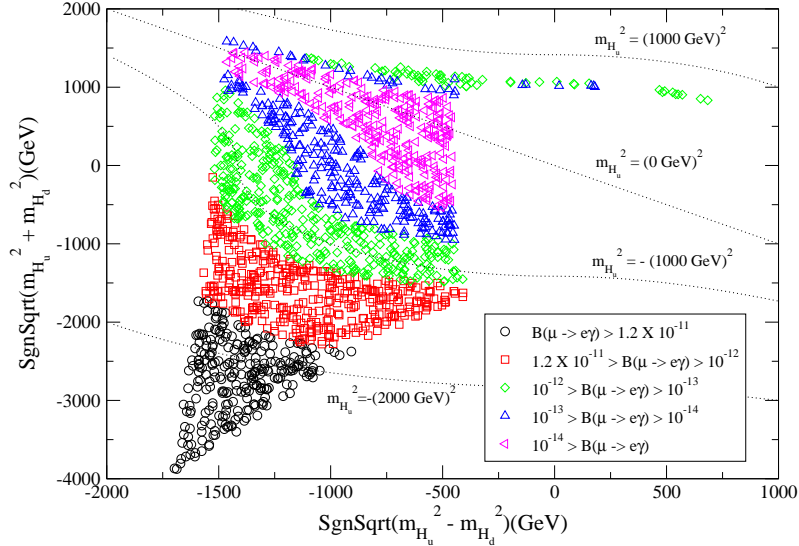}
\vspace{0.3cm} \caption{\label{BRm500b10} $B(\mu\rightarrow
e\gamma)$ as a function of the HENS model parameters $m_{H_u}^2$
and $m_{H_d}^2$. The other model parameters are $M_{1/2}=500\gev$
and $\tan\beta=10$, as well as neutrino-sector parameters
$\theta_{12R}=\theta_{13R}=\theta_{23R}=\pi/4+i\ln(\sqrt{2})$,
$M_{N_3}=10^{12}$ GeV, $M_{N_2}=10^{11}$ GeV, and
$M_{N_1}=10^{10}$. All points in this plot are consistent with
collider phenomenology constraints and have a neutralino LSP. }
\end{center}
\end{figure}

  Fig.~\ref{BRm500b10} shows the dependence of the LFV branching
fraction $B(\mu\to e\gamma)$ on $m_{H_u}^2$ and $m_{H_d}^2$ for
the same neutrino sector parameters as Fig.~\ref{BRm300b10}, but
now with $M_{1/2} = 500\,\gev$.  Also as before, $\tan\beta = 10$,
$sgn(\mu) = 1$, and all points shown are consistent with collider
constraints and have a neutralino LSP. Compared to Fig.~\ref{BRm300b10},
the LFV rates are considerably lower. This can be understood in terms of
the general scaling of all the superpartner masses with $M_{1/2}$,
and the fact that larger superpartner masses suppress the loop functions
appearing in Eq.~\eqref{bramplitude}.  Aside from this scaling,
the shapes of the contours in the two figures are very similar,
with the dominant variation in the branching fraction due
to the changing input value of $m_{H_u}^2$.

\begin{figure}[ttt]
\begin{center}
\includegraphics[width=0.7\textwidth]{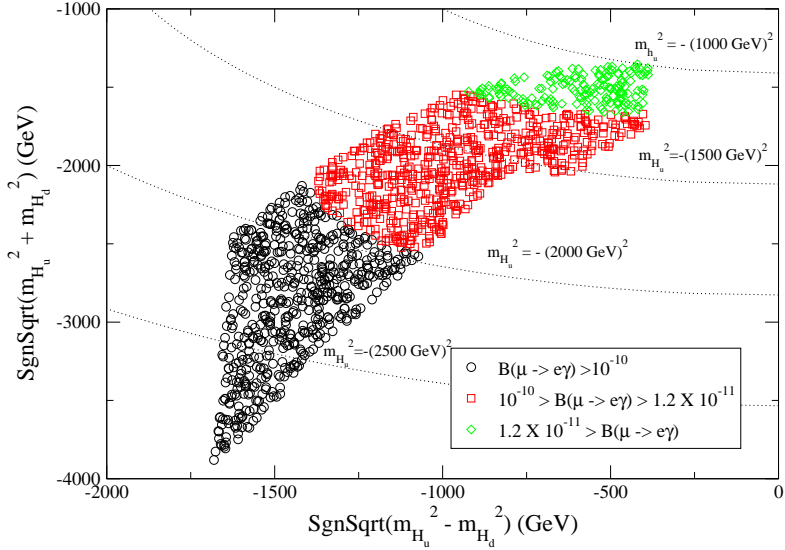}
\vspace{0.3cm} \caption{\label{BRm500b30} $B(\mu\rightarrow
e\gamma)$ as a function of the HENS model parameters $m_{H_u}^2$
and $m_{H_d}^2$. The other model parameters are $M_{1/2}=500\gev$
and $\tan\beta=30$, as well as neutrino-sector parameters
$\theta_{12R}=\theta_{13R}=\theta_{23R}=\pi/4+i\ln(\sqrt{2})$,
$M_{N_3}=10^{12}$ GeV, $M_{N_2}=10^{11}$ GeV, and
$M_{N_1}=10^{10}$. All points in this plot are consistent with
collider phenomenology constraints and have a neutralino LSP. }
\end{center}
\end{figure}

  In Fig.~\ref{BRm500b30} we illustrate the dependence of the
LFV branching ratio $B(\mu\to e\gamma)$ on $m_{H_u}^2$ and
$m_{H_d}^2$ for $\tan\beta = 30$, $M_{1/2} = 500\,\gev$, and $sgn(\mu) = 1$
over the allowed parameter space in the HENS model.
All points in the plot satisfy collider phenomenology
constraints and have a neutralino LSP.
The values of the neutrino sector parameters are the same as in
Figs.~\ref{BRm300b10} and~\ref{BRm500b10}.
The variation of $B(\mu\to e \gamma)$ in
this plot again tracks the value of $m_{H_u}^2$.
However, the overall values of the LFV branching ratio
$B(\mu\to e\gamma)$ are larger than in the previous figures.
There are two reasons for this.  The first is that the
expression for $B(\mu\to e\gamma)$ scales like $\tan^2\beta$.
The second reason for the relative enhancement in the LFV rates
is that larger values of $\tan\beta$ also enhance the $\tau$ Yukawa
coupling, making it more likely to obtain a stau LSP.
To obtain a neutralino LSP, which we demand as a phenomenological
constraint, $m_{H_u}^2$ must be large in magnitude and negative in sign.
This limits the suppression of $B(\mu\rightarrow e\gamma)$
that occurs in the HENS model as $m_{H_u}^2$ becomes small.
With these two sources of relative enhancement at larger values of $\tan\beta$,
we see that in the present example there are very few parameter
points consistent with the bound on $B(\mu\to e\gamma)$
listed in Eq.~\eqref{lfvconstr1}.

\begin{figure}[ttt]
\begin{center}
\includegraphics[width=0.7\textwidth]{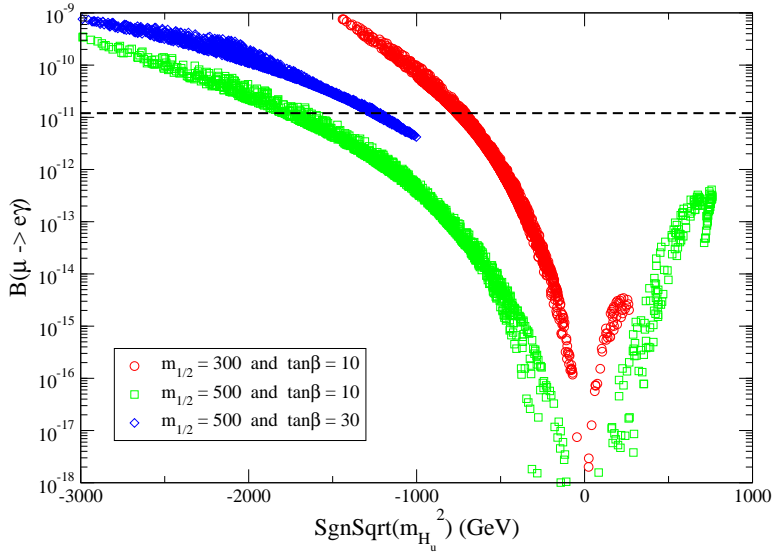}
\vspace{0.3cm} \caption{\label{BRvsMHu} $B(\mu\rightarrow
e\gamma)$ as a function of $m_{H_u}^2$ at the high input scale
for several values of $M_{1/2}$ and $\tan\beta$.
Values of $m_{H_d}^2$ were scanned
over, and all points are consistent with collider phenomenology
constraints and have a neutralino LSP. The neutrino sector
parameters are given by
$\theta_{12R}=\theta_{13R}=\theta_{23R}=\pi/4+i\ln(\sqrt{2})$,
$M_{N_3}=10^{12}\,\gev$, $M_{N_2}=10^{11}\,\gev$, and
$M_{N_1}=10^{10}\,\gev$.
The dashed line in this figure corresponds to the experimental
LFV bound $B(\mu\to e\gamma) < 1.2\times 10^{-11}$.
}
\end{center}
\end{figure}

  In the plots discussed above, the LFV rates depend most
sensitively on the parameter $m_{H_u}^2$.  To better illustrate
this relationship, we plot in Fig.~\ref{BRvsMHu} the same sets of points
as in Figs.~\ref{BRm300b10}, \ref{BRm500b10}, and \ref{BRm500b30}
in terms of $B(\mu\to e\gamma)$ as a function of $m_{H_u}^2$.
These sets correspond to $\tan\beta = 10$ and $M_{1/2} =
300\,\gev$, $\tan\beta = 10$ and $M_{1/2} = 500\,\gev$, and
$\tan\beta = 30$ and $M_{1/2} = 500\,\gev$ respectively,
with $m_{H_d}^2$ scanned over.  The values of the
neutrino sector parameters are the same as in the previous plots.
As expected from Eq.~\eqref{MasIns}, the LFV rates drop precipitously
as $m_{H_u}^2 \to 0$.  When this occurs, only the much smaller
terms beyond the leading order term given in Eq.~\eqref{MasIns}
contribute to lepton flavor mixing.  These subleading terms scale
like $M_{1/2}$, and can not be zeroed out due to the phenomenological
lower bounds on $M_{1/2}$.
Fig.~\ref{BRvsMHu} also illustrates the scaling of
$B(\mu\to e\gamma)$ with $M_{1/2}$, which we expect
to go like $M_{1/2}^{-8}$, as well as the enhancement of the
LFV rates for larger values of $\tan\beta$.
There is a dip in the branching fraction at $m_{H_u}^2\simeq
(700)^2\gev^2$.  This corresponds to $M_1\simeq \mu$, leading
to a large mixing among the neutralinos and a cancellation
between contributions to the amplitude.

  We have concentrated so far on the specific branching fraction
$B(\mu\to e\gamma)$.  The related branching fractions
$B(\tau\to\mu\gamma)$ and $B(\tau\to e\gamma)$ both have a very
similar dependence on the HENS model parameters. Plots of these
branching fractions as a function of $m_{H_u}^2$ are nearly
identical in both shape and overall normalization to those in
Fig.~\ref{BRvsMHu}. However, since the experimental upper bounds
on the branching fractions of these $\tau$ modes are more than a
couple of orders of magnitude larger than the $\mu$ mode, they
provide much weaker constraints on the neutrino-enhanced HENS
parameter space.  We will therefore concentrate most strongly on
the $\mu\to e\gamma$ mode in the present work.

\begin{figure}[ttt]
\begin{center}
\includegraphics[width=0.7\textwidth]{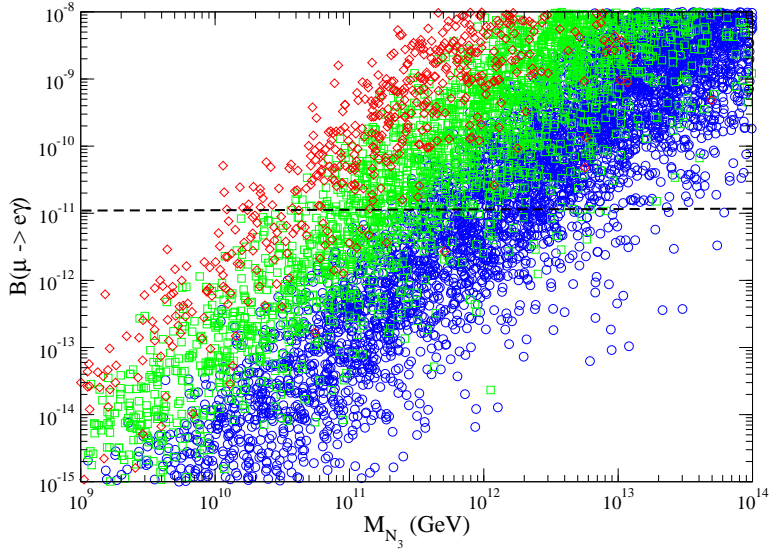}
\vspace{0.3cm} \caption{\label{BRvsMN3}$B(\mu\rightarrow e\gamma)$
as a function of the heaviest right-handed neutrino mass $M_{N_3}$
for the HENS parameters $m_{H_u}^2=-(668)^2\gev^2$,
$m_{H_d}^2=-(511)^2\gev^2$, $\tan\beta=10$, and
$M_{1/2}=300\,\gev$. The blue circles, green squares, and red
diamonds correspond to $Max\{|R|\}<2$, $2<Max\{|R|\}<5$, and
$5<Max\{|R|\}<10$, respectively.  The dashed line represents the
experimental bound of $B(\mu\to e\gamma)<1.2\times 10^{-11}$.}
\end{center}
\end{figure}

  Having studied the dependence of the LFV rates on the HENS model
parameters for a particular (but typical) set of neutrino sector
parameters, let us next examine the dependence of the LFV rates on
the details of the neutrino sector.  In Fig.~\ref{BRvsMN3} we show
the branching fraction $B(\mu\to e\gamma)$ as a function of the
heaviest right-handed neutrino mass $M_{N_3}$.  Of the heavy
neutrino masses, this one usually plays the most important role
in determining the amount of LFV.  The HENS model parameters for
this plot are $\tan\beta = 10$, $M_{1/2} = 300\,\gev$, $m_{H_u}^2
= -(511\,\gev)^2$ and $m_{H_d}^2 = -(668\,\gev)^2$.  These values
produce a phenomenologically consistent spectrum, which we list in
Appendix~B, and are not unusual in terms of LFV. The light
neutrino masses are as described in Appendix~A.  The remaining
neutrino sector parameters were scanned over:  heavy neutrino
masses lie in the range $M_N \in [10^7,10^{14}]\,\gev$ with no particular
hierarchy between them, and the $R$ matrix angles range over
$\textbf{Re}(\theta) \in [0,2\pi]$ and $\textbf{Im}(\theta) \in [-2,2]$.
Within the plot, the blue circles, green squares,
and red diamonds correspond to $Max\{|R|\} \in [0,2]$, $Max\{|R|\} \in [2,5]$,
and $Max\{|R|\} \in [5,10]$. Recall that since $R$ is a complex
orthogonal matrix, its components are unbounded, although large
components require a fine-tuning to obtain small neutrino masses.

  The two most important neutrino sector quantities for
$B(\mu\to e\gamma)$ are the structure of the $R$ matrix
and the value of $M_{N_3}$.  The importance of both quantities
can be seen in Fig.~\ref{BRvsMN3}.
In general, smaller neutrino Yukawa couplings lead to less
lepton flavor mixing. Thus, given Eq.~\eqref{NueYuka},
it is not surprising that smaller components in the $R$ matrix,
and lower values of $M_{N_3}$ lead to lower values of $B(\mu\to e\gamma)$.
What is more interesting is the wide range of values of this branching
fraction for a given fixed value of $M_{N_3}$.  This indicates that
certain textures of the neutrino Yukawa matrix can greatly reduce
the amount of LFV.  On account of these various sensitivities,
it is difficult to demarcate a region of parameter space consistent
with the LFV bounds other than by what we have illustrated
in Fig. \ref{BRvsMN3}.  Certain challenging sets of neutrino
sector parameters require $M_{N_3} < 10^{10}\,\gev$, while
for other neutrino parameters the requirement can be weakened
to $M_{N_3} < 10^{13}\,\gev$.
More concrete constraints can be derived in certain limits,
such as when the right-handed neutrinos are strongly hierarchical.

\section{Leptogenesis in $\nu$HENS\label{TheLep}}

  The primary motivation for heavy right-handed neutrinos
is to explain the findings of neutrino oscillation experiments.
However, heavy neutrinos also provide a mechanism to account
for the baryon asymmetry, which is measured to be~\cite{Komatsu:2008hk}
\begin{equation}
Y_B=\frac{n_B-n_{\overline{B}}}{s}=(8.7\pm 0.3)\times 10^{-11}.
\end{equation}
With heavy right-handed neutrinos, this baryon asymmetry can be
generated through the process of leptogenesis~\cite{Fukugita:1986hr,
Luty:1992un}.  As the universe cools, the heavy neutrinos fall
out of equilibrium and decay.  If there is a significant amount of
CP violation in the neutrino sector, these decays can induce
a net lepton number. This lepton number is subsequently reprocessed
into a net baryon asymmetry through the
$(B+L)$-violating sphaleron transitions~\cite{Kuzmin:1985mm}.
In the present section we investigate whether the HENS model with
heavy right-handed neutrinos can explain the baryon asymmetry
through \emph{thermal} leptogenesis.

  We use the results of Ref.~\cite{Abada:2006ea} to compute the
baryon density due to thermal leptogenesis in the HENS model.
In particular, we take into account \emph{flavor effects}~\cite{
Abada:2006ea,Abada:2006fw,Nardi:2005hs,Nardi:2006fx,Nardi:2007jp}
arising from interactions of the charged Yukawa couplings.
Motivated both by the apparent hierarchy of light neutrino
masses and the desire to reduce the amount of washout of
the lepton asymmetry generated by heavy neutrino decays,
we will focus on mildly hierarchical right-handed neutrino masses,
with $M_{N_1} < M_{N_{2,3}}/3$.

  The baryon asymmetry due to thermal leptogenesis can be expressed
in terms of the CP and L asymmetry $\epsilon_{\alpha}$ and
the effective neutrino mass $m_{\alpha}$ for each flavor
${\alpha} = e,\,\mu,\,\tau$, and the \emph{washout parameter} $\eta$.
In the hierarchical limit of $M_{N_1} \ll M_{N_{2,3}}$,
$\epsilon_{\alpha}$ and $m_{\alpha}$ are given by~\cite{Abada:2006ea}
\begin{eqnarray}
\epsilon_\alpha & \simeq & - \frac{3M_{N_1}}{16\pi
v_u^2}\frac{\textbf{Im} \left[\sum_{i,j}
m_{i}^{1/2}m_{j}^{3/2}U^*_{\alpha i}U_{\alpha j}R_{1i}R_{1j}\right]}
{\sum_k m_{k}|R_{1k}|^2}\\
\widetilde{m}_\alpha &\equiv& \frac{|Y_{\nu
1\alpha}|^2v_u^2}{M_{N_1}}=\left| \sum_k
R_{1k}m_k^{1/2}U^*_{\alpha k}\right|^2.
\end{eqnarray}
A simple approximate form for the washout parameter $\eta$
is~\cite{Abada:2006ea}
\begin{equation}
\eta(\widetilde{m}_{\alpha})=\left[
\left(\frac{\widetilde{m}_{\alpha}}{8.25\times
10^{-3}\, {\rm eV}}\right)^{-1} + \left(\frac{0.2\times 10^{-3}
\, {\rm eV}}{\widetilde{m}_{\alpha}}\right)^{-1.16}\right]^{-1}.\label{Eta}
\end{equation}
The two terms in this expression interpolate between the weak
(first term) and strong (second term) washout regimes.  The first
term in Eq.~\eqref{Eta}, corresponding to weak washout, assumes
there is no initial abundance (thermal or otherwise) of
right-handed neutrinos, and that the only sources of right-handed
neutrinos are inverse decays and scattering. This is a
conservative assumption, as the effective value of $\eta$ can be
enhanced if the initial state has a non-vanishing heavy neutrino
density.

  For right-handed neutrino masses in the range
$(1+\tan^2\beta)\,10^{9}\gev\lesssim M_{N_1} \lesssim
(1+\tan^2\beta)\,10^{12}\gev$ the two lepton flavor approximation
is appropriate, and the resulting baryon density is~\cite{Abada:2006ea}
\begin{equation}
Y_B\simeq
-\frac{10}{31g_*}\left[\epsilon_2\eta\left(
\frac{541}{761}\widetilde{m}_2\right)
+\epsilon_{\tau}\eta\left(\frac{494}{761}\widetilde{m}_{\tau}
\right)\right].\label{YB2}
\end{equation}
In this expression, $g_*$ is the usual number of relativistic
degrees of freedom,
$\widetilde{m}_2 \equiv \widetilde{m}_e+\widetilde{m}_{\mu}$ and
$\epsilon_2 \equiv \epsilon_e+\epsilon_{\mu}$.
For lighter right-handed neutrino states, with mass in the range
$(1+\tan^2\beta)\,10^{5}\,\gev\lesssim M_{N_1} \lesssim
(1+\tan^2\beta)\,10^{9}\,\gev$, we must account for all three lepton
flavors.  The appropriate expression for the baryon
asymmetry in this case is~\cite{Abada:2006ea}
\begin{equation}
\label{YB3}
Y_B\simeq
-\frac{10}{31g_*}\left[\epsilon_e\eta\left(\frac{93}{110}\widetilde{m}_e\right)
+\epsilon_{\mu}\eta\left(\frac{19}{30}\widetilde{m}_{\mu}\right)
+\epsilon_{\tau}\eta\left(\frac{19}{30}\widetilde{m}_{\tau}\right)\right].
\end{equation}

\begin{figure}[ttt]
\begin{center}
\includegraphics[width=0.7\textwidth]{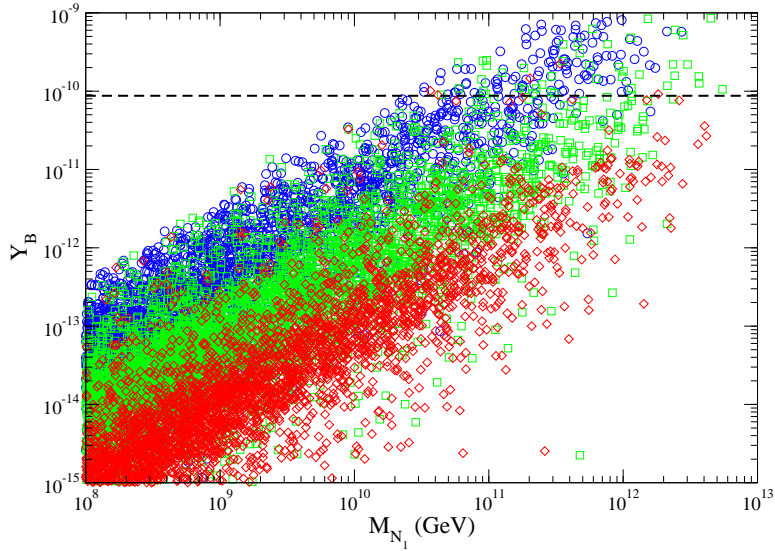}
\vspace{0.5cm} \caption{\label{YBvsMN1} Baryon density due to
leptogenesis in the HENS model as a function of $M_{N_1}$.  The
HENS model parameter were set to $m_{H_u}^2=-(668)^2\gev^2$,
$m_{H_d}^2=-(511)^2\gev^2$, $\tan\beta=10$, and $M_{1/2}=300\,\gev$,
and the neutrino sector parameters were scanned over.   The blue
circles, green squares, and red diamonds correspond to
$Max\{|R|\}<2$, $2<Max\{|R|\}<5$, and $5<Max\{|R|\}<10$,
respectively. The dashed line represents the measured baryon density
$Y_B = (8.7\pm 0.3)\times 10^{-11}$.
}
\end{center}
\end{figure}

  In Fig.~\ref{YBvsMN1} we show the baryon density due to leptogenesis
in the HENS model with heavy right-handed neutrinos as a function
of the lightest heavy neutrino mass $M_{N_1}$.  The HENS model
parameters are set to $M_{1/2} = 300\,\gev$, $\tan\beta = 10$,
$m_{H_u}^2 = -(668\,\gev)^2$, and $m_{H_d}^2 = -(511\,\gev)^2$.
The superpartner mass spectrum for these values is listed in
Appendix~B, and is phenomenologically acceptable
aside from LFV constraints.  We expect these parameters
to be typical in terms of leptogenesis.  The neutrino sector
parameters were scanned over, with the blue circles, green
squares, and red diamonds corresponding to $Max\{|R|\}<2$,
$2 < Max\{|R|\}<5$, and $5<Max\{|R|\}<10$, respectively.

  Fig.~\ref{YBvsMN1} indicates that there is a lower bound on
$M_{N_1}$ if thermal leptogenesis is to be the source of the
baryon asymmetry of the universe.  The minimal value of $M_{N_1}$
that works is on the order of $10^{10}\,\gev$, which is
consistent with the results of
Refs.~\cite{Davidson:2002qv,Giudice:2003jh,Hambye:2003rt}. This
plot also shows  that the final baryon asymmetry is reduced as the
magnitudes of the entries in the $R$ matrix become larger. The
reason for this is that larger values of $|R_{ij}|$ enhance
$\tilde{m}_{\alpha}$, which increases the amount of washout.
In the strong washout regime, which we find to be the case throughout
much of the parameter space, the lepton asymmetry produced in
right-handed neutrino decays is thereby greatly diluted.
To obtain a sufficiently large lepton asymmetry to explain the
baryon excess in this regime, $M_{N_1}$ must be larger than
about $10^{10}\gev$.  This can cause difficulties for avoiding
the experimental constraints on LFV, as we will discuss later.

  Let us also make note of the fact that the lower bound
on $M_{N_1}$ of about $10^{10}\,\gev$ suggests that the reheating
temperature after inflation was larger than this if thermal
leptogenesis is to explain the baryon asymmetry.  In
supersymmetric models, such large reheating temperatures lead to
the overproduction of gravitinos~\cite{Bolz:2000fu}. Within the
HENS model with an input scale on the order of $M_{GUT}$ and an
underlying gravity or gaugino mediation of supersymmetry breaking,
we expect gravitino masses on the order of the weak
scale~\cite{Buchmuller:2005rt}. Gravitinos of this mass decay
during nucleosynthesis, and can ruin the ratios of the light
element abundances for $T_{RH} \gtrsim 10^{7\pm
1}\,\gev$~\cite{Kawasaki:2004qu}. A couple of possible approaches
to this problem are resonant enhancements of the lepton asymmetry
as the heavy neutrinos become nearly degenerate that allow
$M_{N_1}$ to be lowered further~\cite{
Flanz:1994yx,Covi:1996wh,Pilaftsis:1997jf,Pilaftsis:2003gt,Hambye:2004jf},
or the non-thermal production of heavy right-handed neutrinos
after inflation~\cite{Asaka:1999yd,Giudice:1999fb}.

\section{HENS Leptogenesis with LFV Constraints\label{LepLFV}}

  In the previous two sections we have examined the LFV constraints
on the HENS model with heavy right-handed neutrinos,
and we have investigated whether this model can account
for the baryon asymmetry through thermal leptogenesis.
In the present section, we combine these considerations,
and study whether the HENS model can be successful in both
regards at the same time.  To be concrete, we focus on two
particular points in the HENS parameter space.  For these points,
we study many different structures of the neutrino sector, with the
one simplifying assumption of slightly hierarchical right-handed
neutrino masses with $M_{N_1} \lesssim M_{N_{2,3}}/3$.

  We will refer to the two HENS model parameter sets
as points~A and~B.  Both points have $\tan\beta = 10$,
$M_{1/2} = 300\,\gev$, and $sgn(\mu)>0$.  For point~A,
the Higgs sector parameters at the input scale are
$m_{H_u}^2=-(668)^2\gev^2$ and $m_{H_d}^2=-(511)^2\gev^2$.
The corresponding input values for point~B are
$m_{H_u}^2=-(100)^2\gev^2$, $m_{H_d}^2=-(359)^2\gev^2$.
The resulting low-energy spectra for these two points
are phenomenologically consistent, aside from LFV constraints.
We list their mass spectra in Appendix~B.  The crucial difference
between the two parameter points is that the input value of $m_{H_u}^2$
is much larger for point A than for point B.

\subsection{Simultaneous Constraints}

\begin{figure}[ttt]
\begin{center}
\subfigure[]{
\includegraphics[width=0.45\textwidth]{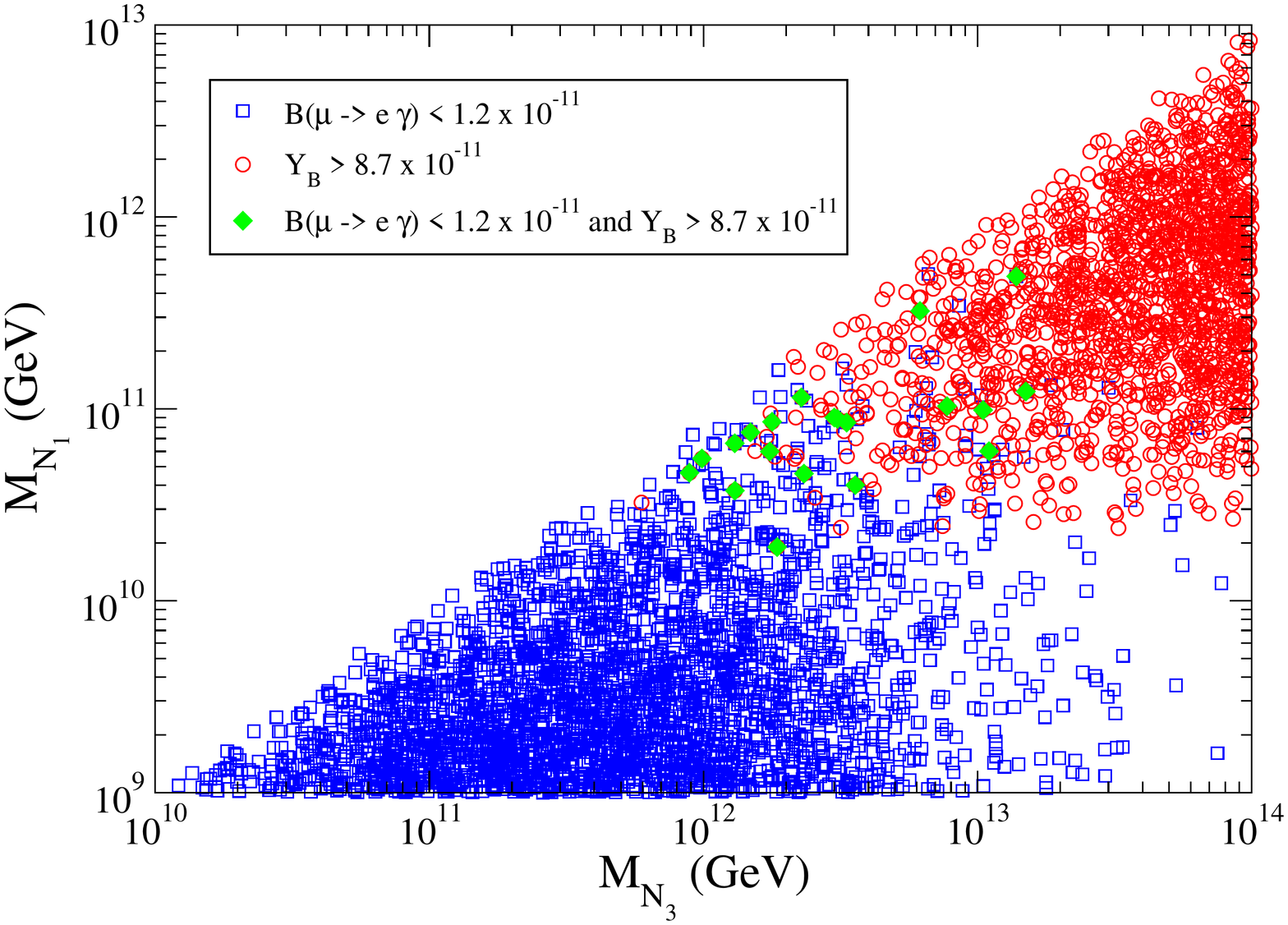}}
\subfigure[]{\includegraphics[width=0.45\textwidth]{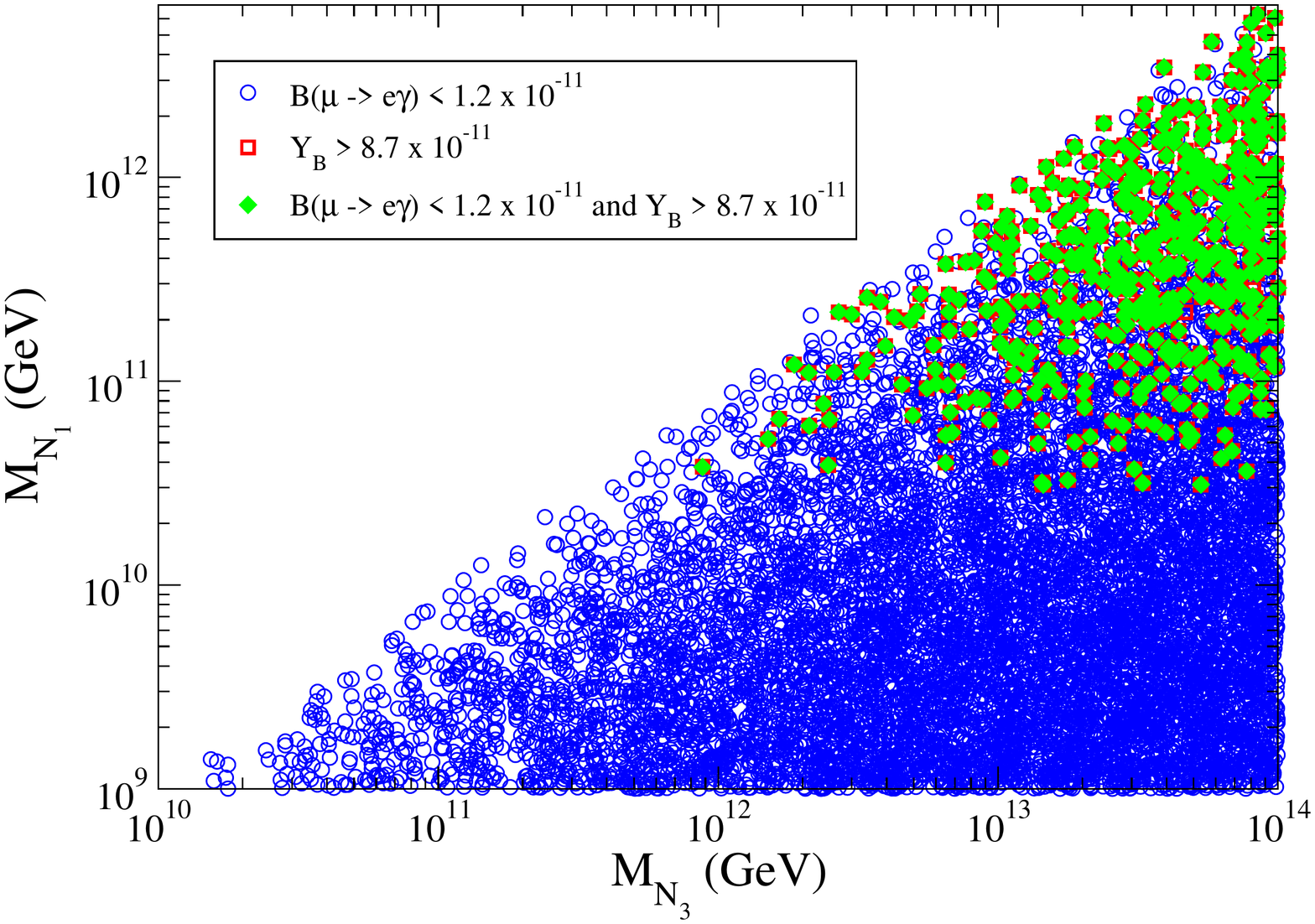}}
\caption{\label{YBBR}  HENS parameter points in the
$M_{N_3}$-$M_{N_1}$ plane consistent with LFV constraints (blue
squares), baryogenesis through thermal leptogenesis (red circles),
or both simultaneously (green diamonds).  The panel on the left
(a) is for HENS parameter set A, with $M_{1/2}=300\gev$,
$\tan\beta=10$, $m_{H_u}^2=-(668)^2\gev^2$, and
$m_{H_d}^2=-(511)^2\gev^2$.  The panel on the right (b) is for
HENS parameter set B, with $M_{1/2}=300\gev$, $\tan\beta=10$,
$m_{H_u}^2=-(100)^2\gev^2$, and $m_{H_d}^2=-(359)^2\gev^2$. In
both plots we have scanned over neutrino sector parameters. }
\end{center}
\end{figure}

  In Section~\ref{LFVHN} we found that LFV constraints
favor smaller values of $M_{N_3}$.  On the other hand, in
Section~\ref{TheLep} we found that thermal leptogenesis prefers
larger values of $M_{N_1} < M_{N_3}$.  The tension between these
two requirements is illustrated in Fig.~\ref{YBBR}, where we plot
points in the $M_{N_3}$-$M_{N_1}$ plane that are consistent with
LFV constraints, that generate enough of a baryon asymmetry
through thermal leptogenesis, or that satisfy both conditions.
The left-hand panel of this figure corresponds to point~A described above,
while the right-hand panel corresponds to point~B. In both panels,
we have scanned over heavy neutrino masses $M_{N_i}$,
as well as the light neutrino masses and the values of the
$U$ and $R$ mixing matrices subject to the constraints listed
in Appendix~A.
The blue squares in the figure are points that obey the LFV
constraints, the red circles are points that generate enough
of a baryon excess, and the green diamonds satisfy both requirements.

  Only a very small subset of the points in Fig.~\ref{YBBR}
for set~A are consistent with both the LFV constraints and
leptogenesis. This is primarily the result of the large value of
$m_{H_u}^2$ for this parameter set, which leads to large LFV rates
unless $M_{N_3}$ is very small. This in turn pushes down the
possible range of values of $M_{N_1}$, making leptogenesis less
effective.  Only for a small and special subset of the neutrino
sector parameters can both requirements be met.  We will discuss
these requirements in more detail below. In contrast, there are
many points for parameter set~B for which both the LFV and
leptogenesis constraints are met. Indeed, very few of the points
that are consistent with generating the baryon asymmetry through
leptogenesis do not satisfy the LFV constraints. This is due to
the LFV constraints being very weak given the relatively small
value of $m_{H_u}^2$ for this parameter set.

\subsection{Neutrino Yukawa Matrix Structures\label{StruYuk}}

  We found above that only a small subset of the neutrino sector
parameters allowed for the HENS parameter set~A to be
consistent with the constraints from LFV while generating the baryon
asymmetry via thermal leptogenesis.  The combination of these
two requirements selects a particular structure for the neutrino
Yukawa matrix which we describe here.
Due to the assumed hierarchy among the right-handed neutrinos,
the Yukawa matrix will generally decrease in size from row three to row one.
Thus, the leading contributions to the off-diagonal
components of $m_{L_{ij}}^2$ responsible for LFV are typically
\begin{equation}
m_{\tilde{L}_{ij}}^2=-\frac{m_{H_u}^2}{8\pi^2}\left(Y_{\nu 3i}^*Y_{\nu
3j}\,t_3+Y_{\nu 2i}^*Y_{\nu
2j}\,t_2\right),\label{SlepYuka}
\end{equation}
where $t_i=\ln(M_{GUT}/M_{N_i})$.
This feature selects out the $Y_{\nu 3i}$ and $Y_{\nu 2i}$ components
of the neutrino Yukawa matrix as being particularly important.

\begin{figure}[ttt]
\begin{center}
\includegraphics[width=0.7\textwidth]{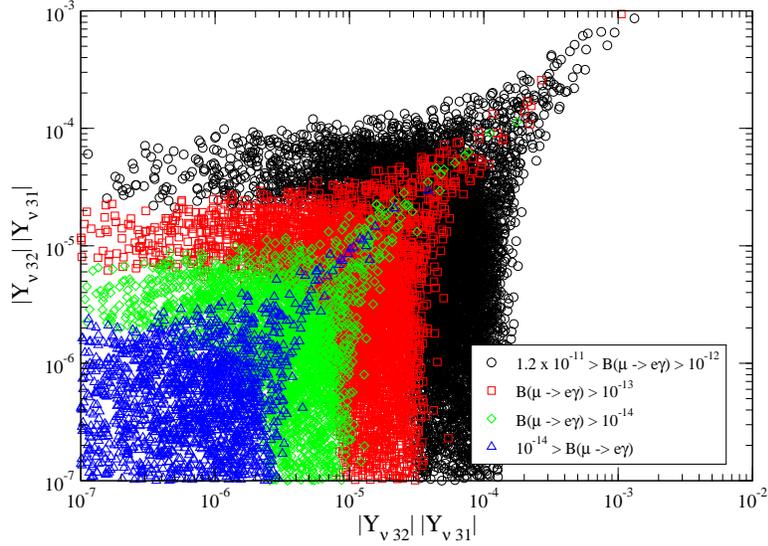}
\vspace{0.3cm}
\caption{\label{YukConsMeG}
$B(\mu\rightarrow e\gamma)$ in the plane of $|Y_{\nu 32}| |Y_{\nu
31}|$ and $|Y_{\nu 22}| |Y_{\nu 21}|$ for the mass spectrum~A in
Appendix~B, corresponding to HENS parameters
$m_{H_u}^2=-(668)^2\gev^2$, $m_{H_d}^2=-(511)^2\gev^2$,
$\tan\beta=10$, and $M_{1/2}=300\gev$.}
\end{center}
\end{figure}

  In Fig.~\ref{YukConsMeG} we show the dependence of the
LFV branching fraction $B(\mu\to e\gamma)$ on the
$Y_{\nu 3i}$ and $Y_{\nu 2i}$ components of the neutrino Yukawa matrix
for the HENS model parameter set~A described above and in Appendix~B.
The points in this plot correspond to different values of the $R$ and
$U$ matrix elements, and (hierarchical) right-handed neutrino masses.
With the spectrum of parameter set~A, the $\mu\to e\gamma$
branching fraction can be written as
\begin{equation}
B(\mu\rightarrow e \gamma) = (1400\,\gev)^{-4}\,|m_{\tilde{L}_{21}}^2|^2.
\end{equation}
With $M_{N_3}= 10^{11}\gev$, for example, this translates into
a constraint on the Yukawa couplings of
\begin{equation}
Y_{\nu 32}^*Y_{\nu 31} + Y_{\nu 22}^*Y_{\nu
21}\frac{t_2}{t_3}<9.6\times 10^{-5}\label{NeuYukBou}
\end{equation}
where $t_i=\ln(M_{GUT}/M_{N_i})$. This constraint can be met in
two different ways: both $|Y_{\nu 32}||Y_{\nu 31}|$ and $|Y_{\nu
22}| |Y_{\nu 21}|$ can be separately very small, or $Y_{\nu
32}^*Y_{\nu 31}$ and $Y_{\nu 22}^*Y_{\nu 21}$ can cancel against
each other.  It is this cancellation that leads to the pointed
structure in Fig.~\ref{YukConsMeG}.

\begin{figure}[ttt]
\begin{center}
\includegraphics[width=0.7\textwidth]{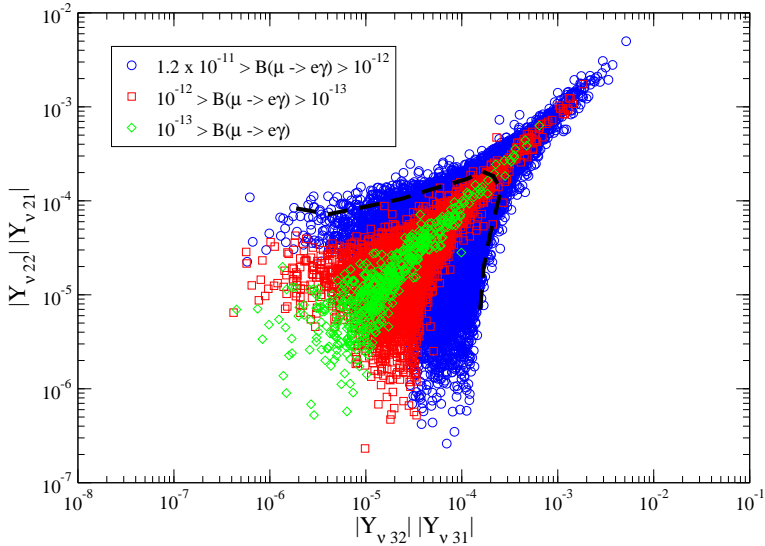}
\caption{\label{YnuYnu}
$B(\mu\rightarrow e\gamma)$ in the plane of $|Y_{\nu 32}| |Y_{\nu
31}|$ and $|Y_{\nu 22}| |Y_{\nu 21}|$ for the mass spectrum~A in
Appendix~B.  All points in this figure can account for the baryon
asymmetry through thermal leptogenesis.
The dashed line corresponds to the region that would still be
allowed if the bound on $\tau\to\mu\gamma$ decay were improved to
$B(\tau\to\mu\gamma) < 10^{-10}$.
}
\end{center}
\end{figure}

  The constraints on the neutrino Yukawa couplings become
even stronger when we demand successful leptogenesis as well.
In Fig.~\ref{YnuYnu} we show the equivalent plot to
Fig.~\ref{YukConsMeG} for HENS parameter set~A, but now restricted
to points that are consistent with thermal leptogenesis.  Clearly,
larger values of the Yukawa couplings are required for successful
leptogenesis. For these points to also be consistent with LFV
constraints, there must be a significant cancellation between
$Y_{\nu 32}^*Y_{\nu 31}$ and $Y_{\nu 22}^*Y_{\nu 21}$ to suppress
$B(\mu\to e\gamma)$, as suggested by Eq.~\eqref{NeuYukBou}.
With the present sensitivities, the bounds on LFV in $\tau$
decays do not significantly constrain the allowed parameter
space in this example.  However, improved sensitivities from
current and future experiments could change this.  To illustrate
the effects of improved experimental bounds, we also draw
a dashed contour in Fig.~\ref{YnuYnu} corresponding
to the parameter region that would be allowed with the
stronger constraint $B(\tau\to \mu\gamma) < 10^{-10}$.
Such a level of sensitivity could potentially be achieved
by super $B$ factories~\cite{Hewett:2004tv}.
The primary effect of an improvement in the $\tau$ sector bounds
is to push the neutrino Yukawa couplings to smaller overall
values.  Improving $B(\mu\to e\gamma)$, on the other hand,
forces more fine tuning among the different neutrino Yukawa
matrix elements.

\begin{figure}[ttt]
\begin{center}
\includegraphics[width=0.7\textwidth]{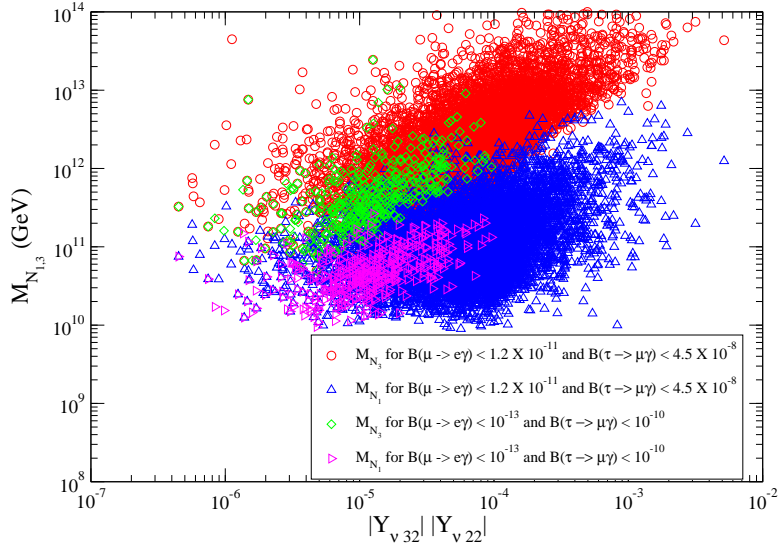}
\caption{\label{YnuYnuMNtmg} Allowed points subject to
the constraints of LFV and thermal leptogenesis for
the HENS model parameter set A.  The points are plotted
as $|Y_{\nu\,32}||Y_{\nu 31}|$ against either $M_{N_1}$
or $M_{N_3}$, with the neutrino sector parameters scanned over.}
\end{center}
\end{figure}

  In Fig.~\ref{YnuYnuMNtmg} we show the allowed regions in
the $|Y_{\nu\,32}||Y_{\nu\,31}|$ and $M_{N_{1,3}}$ planes
for HENS parameter set A points requiring both consistency
with the current LFV bounds as well as successful thermal
leptogenesis.  We have scanned over the neutrino sector
parameters in the same way as in Fig.~\ref{YnuYnu}.
In this plot we also show the regions of the parameter space that
would be allowed if the bounds on LFV were improved to
$B(\mu\to e\gamma) < 10^{-13}$ and $B(\tau\to\mu\gamma)<10^{-10}$.
As discussed above, strengthening the LFV bounds tends to push the
allowed range of $M_{N_3}$ to lower values making leptogenesis less effective.

\section{Conclusions\label{concl}}

  We have investigated the consequences of adding right-handed
neutrinos to the HENS model.  This model provides a simple
and phenomenologically consistent solution to the supersymmetric
flavor problem.  Adding heavy right-handed neutrinos, lepton flavor mixing
can arise due to the neutrino Yukawa matrix in the course
of RG running.  We have studied the constraints on the neutrino-extended
HENS model that arise from the current bounds on LFV.  We have also
investigated whether the baryon asymmetry can be explained
by thermal leptogenesis induced by the heavy right-handed neutrinos.

  We find that the neutrino-extended HENS model can be consistent with
the existing bounds on LFV in two ways.  First, the neutrino
Yukawa couplings that contribute to lepton flavor mixing can be
very small.  In the context of a seesaw generating the light
neutrino masses, this corresponds to lower values of the
right-handed neutrino masses, below about $10^{11}\,\gev$.  The
second way to suppress LFV in the HENS model to arrange for
$m_{H_u}^2$ to be small at the input scale $M_{GUT}$.  It is this
soft mass that combines with the neutrino Yukawa couplings to
source flavor mixing in the RG running.  Taking $m_{H_u}^2\to 0$
therefore strongly suppresses LFV, even for larger values of the
heavy neutrino masses.

  In models with heavy right-handed neutrinos,
the baryon asymmetry of the universe can be successfully explained
by (thermal) leptogenesis.  For this mechanism to be effective
in the HENS model, the mass of the lightest right-handed neutrino
must exceed about $10^{10}\,\gev$.  This implies a tension
with the constraints from LFV.  For both requirements to be met,
either $m_{H_u}^2$ must be somewhat small or the neutrino Yukawa
matrix must have a special structure.
These constraints will be strengthened by current and upcoming
searches for lepton flavor violation.

  Our focus has been on enabling a theoretical idea (HENS) to be
compatible with additional phenomenological requirements (neutrino
masses and small LFV) and explanatory opportunities (baryon
asymmetry). Throughout this work, however, it should be noted that
even though the HENS idea started out by minimizing LFV in
low-scale supersymmetric theories, full compatibility with nature
reintroduced flavor violations through neutrino Yukawa effects.
This is a generic feature of supersymmetric theories that
explicitly incorporate neutrino masses in the spectrum. As
explained above, we find LFV bounds nontrivial to satisfy if the
baryon asymmetry of the universe originates from thermal
leptogenesis with hierarchical right-handed neutrinos. In our
view, this highlights in yet another context the importance of
making progress in LFV experiments whose non-zero signal upon
reaching better sensitivity will be complementary to the knowledge
gained from high-energy LHC experiments and will be necessary to
unravel the underlying theory.

\section*{Acknowledgements}
  We would like to thank Aaron Pierce and Krzysztof Turzynski for
helpful conversations.  This work was supported by the DOE, the
Michigan Center for
Theoretical Physics~(MCTP), the Korean Institute for Advanced Study~(KIAS),
and the Kavli Institute for Theoretical Physics~(KITP) under
the National Science Foundation Grant No. PHY05-51164.


\appendix

\section*{Appendix A: Light Neutrino Parameters\label{lneutstuff}}

  Neutrino experiments have measured the value of two independent
neutrino mass differences: the solar neutrino mass,
$\Delta m_{\odot}^2$, and the atmospheric neutrino mass,  $\Delta
m_{@}^2$.  The $2\sigma$ ranges of these
mass differences are~\cite{Mohapatra:2005wg}
\begin{eqnarray}
\Delta m_{@}^2 &=&|m_{\nu_3}^2-m_{\nu_2}^2| =
(2.1-2.7)\times 10^{-3} \mbox{eV}^2\\
\Delta m_{\odot}^2 &=& m_{\nu_2}^2-m_{\nu_1}^2 =
(7.3-8.1)\times 10^{-5}
\mbox{eV}^2.
\end{eqnarray}
Since the sign of the atmospheric mass difference is
undetermined, the hierarchy of the neutrino masses is unknown.

  With two known mass differences and three light neutrinos,
we can parametrize the masses of all three neutrinos in terms
of a single parameter $m_3$.  In the case of a normal hierarchy~(NH),
we have
\beq
m_3=m_3,~~~
m_2=\sqrt{m_3^2-\Delta m_{@}^2},~~~
m_1=\sqrt{m_3^2-\Delta m_{@}^2-\Delta m_{\odot}^2}.
\eeq
Demanding that the mass of the lightest right-handed neutrino
be real, we obtain a lower bound on the heaviest
left-handed neutrino of
\begin{equation}
m_3=\sqrt{\Delta m_{@}^2 + \Delta m_{\odot}^2}\simeq (0.047\!-\!0.053)
\mbox{eV}.
\end{equation}
We focus on the normal hierarchy in the present work, but we
expect our results will be qualitatively the same for an
inverted hierarchy~(IH).

  Whenever we fix a set of low energy neutrino parameters in our analysis,
we consider the normal hierarchy with neutrino masses of
\beq
m_1=9.0\times 10^{-4}\, {\rm eV},~~~
m_2=9.0\times 10^{-3}\, {\rm eV},~~~
m_3=5.0\times 10^{-2}\, {\rm eV}.\label{NeMas}
\eeq
For the mixing angles in the $U$-matrix, defined in Eq.~\eqref{Umat},
we use the central values of $\theta_{12}$ and $\theta_{23}$,
and set $\theta_{13}=0$.
\beq
\theta_{12}=35^{\circ},~~~
\theta_{13}=0^{\circ},~~~
\theta_{23}=45^{\circ}.
\eeq
These light neutrino parameters are the low-scale values.
We do not consider additional RG running of the light neutrino masses.
As shown in \cite{Antusch:2003kp}, the RG effects will only make
a difference of $10-15\%$.  This will not qualitatively alter our results.


\section*{Appendix B: Sample Mass Spectrum\label{samplemass}}

  We list in Table~\ref{parameters} the high-scale input HENS model
parameters for points~A and~B discussed in the text.
We also list some of the relevant low-scale model parameters
obtained by RG running using SuSpect\,2.34~\cite{Djouadi:2002ze}.
In Table~\ref{masses} we collect the relevant superpartner masses
corresponding to points~A and~B.  Again, these were obtained
using SuSpect\,2.34~\cite{Djouadi:2002ze}.

\begin{table}[htt]
\begin{center}
\begin{tabular}{|c|c||c|c||c|c||c|c|}
\hline
Parameter&A (GeV)& B (GeV)\\
\hline
\hline
$M_{1/2}$&300&300\\
$\tan\beta$&10&10\\
$SgnSqrt(m_{H_u}^2)$&$-668$ & $-100$\\
$SgnSqrt(m_{H_d}^2)$&$-511$ & $-359$\\
$sgn(\mu)$& +&+\\
\hline
\hline
$M_1$ &123&122\\
$M_2$ &231&230\\
$\mu$ & 666&401\\
$M_{A^0}$ & 851&566\\
$m_{\tilde{L}_{1,2}}^2$ &148 &192\\
$m_{\tilde{E}_{1,2}}^2$ &221&140 \\
\hline
\end{tabular}
\end{center}
\caption{High-scale HENS model input parameters and selected
low-scale output parameters for the sample points~A and~B
discussed in the text.\label{parameters}}
\end{table}

\begin{table}[hbb]
\begin{center}
\begin{tabular}{|c|c|c||c|c|c|}
\hline
Particle&A~(GeV)&B~(GeV)&Particle&A~(GeV)&B~(GeV)\\
\hline
\hline
$m_{\chi_1^0}$& 120&118&$m_{\tilde{\nu}_e}$& 134 &180\\
$m_{\chi_2^0}$& 230&219&$m_{\tilde{e}_L}$& 155&197\\
$m_{\chi_3^{0}}$&  667&407&$m_{\tilde{e}_R}$& 225&146\\
$m_{\chi_4^{0}}$& 673 & 425&$m_{\tilde{\nu}_{\tau}}$&  131&179\\
$m_{\chi_1^{\pm}}$& 230&219&$m_{\tilde{\tau}_1}$&  136&132\\
$m_{\chi_2^{\pm}}$& 674&425&$m_{\tilde{\tau}_2}$& 231&201\\
\hline
\end{tabular}
\end{center}
\caption{Low-scale superpartner masses
for the sample points~A and~B discussed in the text.\label{masses}}
\end{table}

\clearpage


\end{document}